\documentclass[pra,amsmath,twocolumn,showpacs]{revtex4}
\usepackage{bm}
\usepackage{amssymb}
\usepackage{graphicx}

\begin{document}
\def\ra{{\rangle}}
\def\la{{\langle}}
\def\^+{{^\dagger}}
\def\n{{\newline}}
\def\Epsilon{{\mathcal{E}}}

\title{Unusual decoherence in qubit measurements with a Bose-Einstein condensate}

\author{D. Sokolovski}
 \affiliation{School of Mathematics and Physics, Queen's University of Belfast,
Belfast BT7 1NN, UK}

\email{d.sokolovski@qub.ac.uk}

\author{S.A. Gurvitz}
 \affiliation{Department of Particle Physics,
 Weizmann Institute of Science, Rehovot 76100, Israel
 }

\email{shmuel.gurvitz@weizmann.ac.il}

\date{\today}


\begin{abstract}
We consider  an electrostatic qubit located near a Bose-Einstein
condensate (BEC) of noninteracting bosons in a double-well
potential, which is used for qubit measurements. Tracing out the BEC
variables we obtain a simple analytical expression for the qubit's
density-matrix. The qubit's evolution exhibits a slow
($\propto1/\sqrt{t}$) damping of the qubit's coherence term, which
however turns to be a Gaussian one in the case of static qubit. This
stays in contrast to the exponential damping produced by most
classical detectors. The decoherence is, in general, incomplete and
strongly depends on the initial state of the qubit.
\end{abstract}

\maketitle

\section{Introduction}

\begin{figure}
\includegraphics[width=8cm]{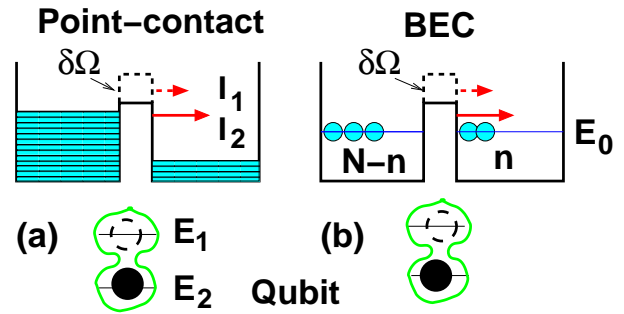}
\caption{(Color on line) a) Point-contact in which electrons,
initially in the left reservoir, tunnel across a potential barrier
modulated by the presence of an electron in one of the two coupled
quantum dots (qubit); b) Atoms of a BE condensate tunnel across a
barrier of a symmetric double-well potential modulated by a qubit. }
\label{fig1}
\end{figure}

Recent progress in quantum information technology has lead to
significant technological and theoretical advances in measuring and
controlling the state of a two-level quantum system (qubit). Devices
used for this purpose include point-contact detectors and single
electron transistors \cite{devor,pc,set} where the magnitude of
electron current is used to determine the qubit's state. Recently,
more sophisticated hybrid systems which combine a charged qubit with
microwave resonators or ensembles cold polar molecules were proposed
\cite{res,hybrid}. In addition to technological benefits such
hybrids offer an insight into fundamental physical phenomena, such
as decoherence. The decoherence is present for any microscopic
system (e.g., a qubit) interacting with a macroscopic device,
characterized by a large number of degrees of freedom and a dense
distribution of energy levels. As a result, an initial state of a
qubit is expected to be rapidly (exponentially in time) converted
into a statistical mixture, so that the information stored in the
qubit is erased \cite{zurek}. For example, this has been explicitly
demonstrated for qubit measurements with a point-contact detector
shown in Fig.\ref{fig1}a, where a macroscopic current flowing into
the right reservoir across the potential barrier is modulated by the
qubit's electron \cite{gur}.

In this letter we study measurements in a hybrid system, consisting
of an electrostatic qubit placed in closed proximity to a
non-interacting BEC trapped in a symmetric double-well potential.
The qubit is represented by an electron in coupled quantum dots
(Fig.\ref{fig1}b), while confinement of the BEC can be realized, for
example, by means of a quasi-electrostatic optical dipole trap
produced by two crossed laser beams \cite{dipole}. Since the
trapping occurs due  the interaction of the induced atomic dipole
moment of neutral atoms and the far-detuned optical field
\cite{roy}, additional electric filed induced by the electron would
change the barrier height. This, in turn, would modulate the atomic
current just as the presence of an electron in one of the dots
modifies the current of the point-contact detector shown in
Fig.\ref{fig1}a. There is, however, an important difference as only
a single level (zero-width band) is available for the tunneling
atoms, which raises the question of what type of decoherence, if
any, would experience the measured qubit?

With the number of carriers macroscopically large, but only one level
existing in each of the reservoirs, the question cannot
be answered without a detailed analysis. On one hand, the bosons are
moving independently, and one could expect their effect to be
similar to that of a single boson which, as is easy to show, does
not produce decoherence. On the other hand, it is not clear whether
the large number of uncorrelated degrees of freedom in the detector
will not have an averaging effect on the qubit thus causing the
off-diagonal elements of its density matrix to disappear.

In the following we will show that the density matrix of a qubit
coupled to a BEC will undergo an evolution which is not be described
by Bloch-like equations similar to those arising in the case of a
point-contact detector \cite{gur}.  However, the BEC model shown in
Fig.\ref{fig1}b can be solved exactly.  We will demonstrate that the
rate of decoherence is extremely (non-exponentially) slow and,
unlike in the case of the point-contact, its amount strongly depends
on the choice of the qubit's initial state.

\section{Description of the model}

Consider an electrostatic qubit interacting with a BEC consisting of
$N$ atoms initially trapped in the left well of a symmetric the
double-well structure (Fig.~\ref{fig1}b). The entire system can be
described by the tunneling Hamiltonian $H=H_{\rm c}+H_{\rm q}+
H_{\rm int}$, where the three terms correspond to the condensate,
the qubit and the interaction, respectively, and
\begin{subequations}
\label{a1}
\begin{eqnarray}
H_{\rm c}&=&E_0(c_L^\dagger c_L+c_R^\dagger c_R)-\Omega (c_L^\dagger
c_R+c_R^\dagger c_L)\label{a1a}\\
H_{\rm q}&=&E_1 d_1^\dagger d_1+E_2 d_2^\dagger d_2-\omega_0
(d_1^\dagger d_2+d_2^\dagger d_1)\label{a1b}\\
H_{\rm int}&=&\delta\Omega\ d_2^\dagger d_2 (c_L^\dagger
c_R+c_R^\dagger c_L).\label{a1c}
\end{eqnarray}\end{subequations} Here $c_{L,R}^\dagger$ is a boson
creation operator in the left (right) reservoir, and
$d_{1,2}^\dagger$ is the fermionic creation operator for the qubit.
Tunneling between the reservoirs is suppressed  when the electron is
in the dot nearest to  the barrier, so that  $\delta\Omega
=\Omega-\Omega'>0$, Fig.\ref{fig1}b.

For a static qubit trapped in one of the quantum dots,
($\omega_0=0$), each boson of the condensate oscillates between the
reservoirs with the Rabi frequency $\Omega$ or $\Omega'$.
Thus, the
probability of finding $n$ bosons in the right reservoir at the time
$t$ is given by
\begin{align}
P_n(t)=\binom{N}{n} \cos^{2(N-n)}(\Omega\, t)\,\sin^{2n}(\Omega\,
t)\, , \label{a0}
\end{align}
if the qubit's electron occupies the level $E_1$ of the nearer dot
(or by the same expression with $\Omega\to\Omega'$ when it occupies
the level $E_2$). In the interesting case when the tunneling rate
for each atom is small but the number of atoms is large, we put
$N\to\infty$, while $\sqrt{N}\,\Omega\to \kappa=$const, thus
maintaining a finite current into the right reservoir. For small
times, $t=\Delta t << \Omega^{-1}$ Eq.~(\ref{a0}) yields $P_n(\Delta
t)\approx (\kappa\, \Delta t)^{2n}\,e^{-(\kappa\, \Delta t)^2}/n! $
and, in particular, $P_0(\Delta t)\approx1- (\kappa\Delta t)^2$.
Such a non-Markovian behavior of the BEC is in contrast with large
fermion reservoirs (Fig.~\ref{fig1}a), where one finds $1-P_0(\Delta
t)\propto (\Delta t)$, which is typical for a Markovian process.

Consider now the behavior of a dynamic qubit, $\omega_0\not =0$,
subjected to a measurement with such a non-Markovian (BEC) detector.
The wave function of the entire system can be written as $|\Psi
(t)\rangle =\sum_{q,n} \psi_{qn}(t)|q\ra |\phi_n\rangle$
 where
\begin{eqnarray}\label{a3}
|q\ra|\phi_n\rangle
\equiv d_q^{\dagger}|0\rangle_{qub}{(c_L^\dagger)^{N-n}(c_R^\dagger)^{n}|0\rangle_{cond}
\over\sqrt{(N-n)!n!}}\\
\nonumber
\end{eqnarray}
is the state corresponding the qubit localized in one of the quantum
dots, $q=1,2$, and $n=0,1,2,\ldots$ is a number of bosons contained
in the right well. For the corresponding probability amplitude
$\psi_{qn}(t)$ we write $\psi_{qn}(t)=\langle q|\la\phi_n|\exp
(-Ht)|q_0\ra|\phi_0\rangle$ state from which the reduced density matrix of
the qubit is obtained by tracing out the BEC states,
\begin{align} \sigma_{qq'}(t)=\sum_{n}
\psi_{qn}(t)\psi_{q'n}^*(t)\ . \label{a4}
\end{align}
Putting, for convenience, $E_0=0$ one easily finds from
Eqs.~(\ref{a1}) that $H_{\rm q}+H_{\rm int}$ commutes with $H_{\rm
c}$ and the evolution operator $U(t)\equiv \exp (-Ht)$ can be
factorized,
\begin{align}
U(t)=e^{-iH_{\rm c}t}e^{-i(H_{\rm q}+H_{\rm int})t} \equiv U_{\rm
c}(t)U_{\rm qint}(t).
\end{align}
Operators $U_{\rm c}(t)$ and $U_{\rm qint}(t)$ can be written in a
simple form in basis of the eigenstates of the BEC Hamiltonian,
\begin{align}
|\tilde\phi_n\rangle ={(c_L^\dagger + c_R^\dagger)^{N-n}(c_L^\dagger
-c_R^\dagger)^{n}|0\rangle\over \sqrt{2^N(N-n)!n!}}\, ,\label{a5}
\end{align}
such that  $H_{\rm c}|\tilde\phi_n\rangle =(N-2n)\Omega
|\tilde\phi_n\rangle$. Indeed, we have $\langle
q,\tilde\phi_n|U_{\rm c}(t)|q',\tilde\phi_{n'}\rangle =\exp [
-i(N-2n)\Omega\, t]\delta_{nn'}\delta_{qq'}$ and also
\begin{align}
\langle q|\la \tilde\phi_n|U_{\rm
qint}(t)|q'\ra|\tilde\phi_{n'}\rangle =\langle q|\hat
U(t,\varepsilon_n )|q'\rangle\delta_{nn'}\, ,\label{a6}
\end{align}
where $\varepsilon_n=(2n-N)\delta\Omega$. It is readily seen that
$\hat U(t,\varepsilon_n)=\exp [-i(H_{\rm q}+\varepsilon_n\,
d_2^\dagger d_2)t]$ represents the evolution operator of an isolated
{\em asymmetric} qubit with the level displacement
$\varepsilon'_n=\varepsilon_n+E_2-E_1$ (asymmetry parameter), whose
matrix elements are easily found to be
\begin{align}
\la 1|\hat{U}(t,\varepsilon_n)|1\ra &= \left[\cos\left(\omega t
\right) -i{\varepsilon'_n\over 2 \omega} \sin\left(\omega
t\right)\right]
e^{-i{\varepsilon'_n t/2}} \nonumber\\
\la 2|\hat{U}(t,\varepsilon_n)|2\ra &=\left[\cos\left(\omega
t\right) +i{\varepsilon'_n\over 2\omega} \sin\left(\omega
t\right)\right]
e^{-i{\varepsilon'_n t/2}}\nonumber\\
\la 1|\hat{U}(t,\varepsilon_n)|2\ra &=-i {\varepsilon'_n\over\omega}
\sin\left(\omega t\right)e^{-i\varepsilon'_n t/2}\label{a7}
\end{align}
where $\omega\equiv\omega
(\varepsilon_n)=\sqrt{(\varepsilon'_n/2)^2+\omega_0^2}$ is the
qubit's Rabi frequency and $\la 2|\hat{U}(t,\varepsilon_n)|1\ra=\la
1|\hat{U}(t,\varepsilon_n)|2\ra$.

For the reduced density matrix of the qubit in Eq.~(\ref{a4}), with
the help of Eqs.~(\ref{a4})-(\ref{a7}) we find
\begin{align}
\sigma_{qq'}(t)=\sum_n
\sigma_{qq'}(t,\varepsilon_n)|\la\phi_0|\tilde\phi_n\ra |^2\,
,\label{a8}
\end{align}
where
\begin{align}
\sigma_{qq'}(t,\varepsilon_n)=\la q|\hat
U(t,\varepsilon_n)|q_0\ra\la q_0|\hat U^{-1}(t,\varepsilon_n)|q'\ra
\, .\label{aa8}
\end{align}
is the density matrix corresponding to the unitary evolution of an
isolated asymmetric qubit. With the initial state of the BEC given
by $|\phi_0\rangle = (1/\sqrt{N!})(c_L^\dagger )^N|0\rangle$ we then
find
\begin{align}
|\la\phi_0|\tilde\phi_n\ra |^2={N!\over 2^Nn!(N-n)!}\simeq
{2\over\sqrt{2\pi N}}e^{-{(N-2n)^2\over 2N}}
 \label{a9}
\end{align}
where we have used the Sterling formula $K!\simeq \sqrt{2\pi
K}K^K\exp (-K)$ to evaluate the factorials.

Now we assume that the qubit's coupling with each individual atom of
the condensate $(\delta\Omega)$ is weak, but its interaction with
the entire condensate is considerable, and so is the variation of
the BEC current ($\propto \sqrt{N}\,\delta \Omega$), induced by the
qubit. Then taking the limit
\begin{align}
N\to\infty, \quad{\mbox{with}}\quad \sqrt{N}\,\delta \Omega\to
\alpha={\mbox{const}} \label{a9a}
\end{align}
we replace  the sum over $n$ in (\ref{a8}) by an integral,
$\sum_n\to \int d\varepsilon/(2\,\delta\Omega )$. This yields
\begin{equation}
\sigma_{qq'}(t)=\int_{-\infty}^{\infty}\tilde\sigma_{qq'}(t,\varepsilon
)\exp (-\varepsilon^2/ 2\alpha^2) {d\varepsilon\over
\sqrt{2\pi}\alpha} \label{a10}
\end{equation}
where $\tilde\sigma_{qq'}(t,\varepsilon )\equiv
[\sigma_{qq'}(t,\varepsilon )+\sigma_{qq'}(t,-\varepsilon )]/2$.

In the following we will consider only the case of a symmetric
qubit, $E_1=E_2$ \cite{symm}. Then for an initial qubit's state
$|q_0\ra =(a\, d_1^\dagger +b\, d_2^\dagger)|0\ra_q$, we obtain from
Eq.~(\ref{aa8})
\begin{subequations}
\label{a11}
\begin{align}
&\tilde\sigma_{11}(t,\varepsilon )=|a|^2+(|b|^2-|a|^2)[1-\cos
(2\omega t)](\omega_0^2/2\omega^2)
\nonumber\\
& -{\rm Im}(ab^*)\sin (2\omega t)(\omega_0/\omega)\, ,
\label{a11a}\\
&\tilde\sigma_{12}(t,\varepsilon )=i(|a|^2-|b|^2)\sin (2\omega
t)(\omega_0 /2\omega )+ ab^*\cos(2\omega t)\nonumber\\& +{\rm
Re}(ab^*)[1-\cos (2 \omega t)](\omega_0/\omega)^2  \, , \label{a11b}
\end{align}
\end{subequations}
with $\omega\equiv\omega(\varepsilon ) =
\sqrt{(\varepsilon/2)^2+\omega_0^2}$) and
$\tilde\sigma_{22}(t)=1-\tilde\sigma_{11}(t)$,
$\tilde\sigma_{21}(t)=\tilde\sigma_{12}^*(t)$.

\section{Decoherence of qubit due to its interaction with the BEC}

The simple form of Eqs.(\ref{a10})-(\ref{a11}) allows for an easy
analysis of limiting cases. Indeed, the strength of interaction with
the BEC, $\alpha$, enters Eq.(\ref{a10}) only via the Gaussian
cut-off factor $\exp(-\varepsilon^2/2\alpha^2)$. The factor
determines the number of asymmetric configurations contributing of
the qubit's evolution and, therefore the perturbation incurred upon
the qubit by the BEC. (Note that when the interaction vanishes,
$\alpha \rightarrow 0$, the Gaussian becomes narrow, and we recover
the unperturbed evolution of the isolated qubit.)

It is readily seen that in the large time limit  $t\to\infty$ the
contributions for the rapidly oscillating terms in
Eqs.(\ref{a11}a)-(\ref{a11b}) vanish. Evaluating the remaining
integrals analytically shows that as $t\to\infty$ the density matrix
of a qubit tends to a steady state $\sigma^{st}$ given by
\begin{align}
&\sigma^{st}_{11}= |a|^2+(\sqrt{\pi}/2)z\,\exp\,(z^2)\,{\rm erfc}\,
(z)\, (|b|^2-|a|^2)\, ,
\nonumber\\
&\sigma^{st}_{12}= \sqrt{\pi}\, z\exp\,(z^2)\, {\rm erfc}\, (z) {\rm
Re}(ab^*)\, ,\label{a12}
\end{align}
where $z=\sqrt{2}\omega_0/\alpha$ and ${\rm
erfc}\,(z)={2\over\sqrt{\pi}} \int_z^\infty \exp(-t^2)dt$ is the complementary
error function.

\begin{figure}
\includegraphics[width=8cm]{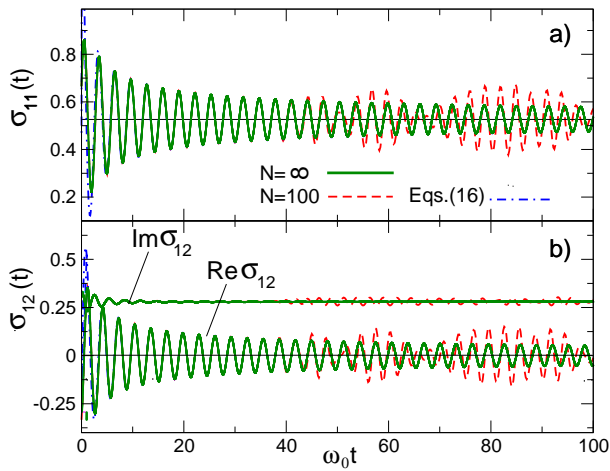}
\caption{(Color on line) a) The qubit's density matrix element
$\sigma_{11}$ vs. time $t$ for $|q_0\ra=
[(1+i)|1\ra+i|2\ra]/\sqrt{3}$, $\alpha=1$, $N\rightarrow \infty$,
Eq.(\ref{a10}), (solid), $N=100$, Eq.(\ref{a8}),  (dashed) and the
stationary phase approximation, Eqs.~(\ref{a13}) (dash-dot).
Horizontal line shows the large time asymptote (\ref{a12}) ; b) same
as (a) but for $\sigma_{12}$.} \label{fig2}
\end{figure}

To evaluate the speed with which this steady state is attained we
note that at large $t$ the phase of the sines and cosines in
Eqs.(\ref{a11}) develops a stationary region of the width $\Delta
\varepsilon = (4\omega_0/t)^{1/2}$ centered at $\varepsilon =0$.
Once $\Delta \varepsilon$ becomes small compared to the width of the
Gaussian in Eq.(\ref{a10}), i.e. for $t >> \omega_0/\alpha^2$, the
contribution from the stationary region becomes proportional to
$\Delta \varepsilon$ causing the time-dependent part of
$\sigma_{qq'}$ in Eqs.(\ref{a11}) to decay as $1/\sqrt{t}$. (For a
recent discussion of non-exponential decoherence behavior expected
in other systems see, for example, Ref.\cite{nexp,nexp1}.)
Explicitly, for $\omega_0\ne 0$, the stationary phase method yields:
\begin{subequations}
\label{a13}
\begin{align}
&\sigma_{11}(t) \simeq \sigma^{st}_{11}+ \sqrt{\omega_0\over
2\alpha^2t}\,
\Big[(|a|^2-|b|^2)\cos\Big(2\omega_0t+{\pi\over4}\Big)\nonumber\\
&~~~~~-2\,{\rm Im}\, (ab^*)\sin \Big(2\omega_0t+{\pi\over4}\Big)\Big]\label{a13a}\\
&\sigma_{12}(t) \simeq \sigma^{st}_{12}+i\sqrt{\omega_0\over
2\alpha^2t}\, \Big[(|a|^2-|b|^2)\sin\Big(2\omega_0t+{\pi\over4}\Big)
\nonumber\\
&~~~~~+2\,{\rm Im}\,
(ab^*)\sin\Big(2\omega_0t+{\pi\over4}\Big)\Big]\, , \label{a13b}
\end{align}
\end{subequations}
where $\sigma^{st}$ is given be Eq.~(\ref{a12}). Figure \ref{fig2}
demonstrates that Eq.(\ref{a13}) (dot-dashed curve) coincides to
graphical accuracy with the exact result (\ref{aa8}) (solid curve)
except at very short times.

Equations (\ref{a12}), (\ref{a13}) which describe the qubit's
decoherence generated by the BEC employed as a measurement device
represent our main result. The qubit's behavior is very different
from that of a qubit interacting with electronic
reservoirs\cite{gur}, Fig.~\ref{fig1}, or in a general with any
Markovian environment, whose effect can be described by Bloch-like
equations \cite{slich,leggett,gur2}. Indeed, it follows from
Eqs.~(\ref{a13}) that the relaxation to the final steady state is
extremely {\it slow}, obeying the power law $\propto 1/\sqrt{t}$.
One exception from this rule is a static qubit ($\omega_0=0$) for
which the stationary region vanishes so that from Eqs.~(\ref{a10}),
(\ref{a11}), one easily obtains $\sigma_{12}(t)=ab^* \exp
(-\alpha^2t^2/2)$. In contrast, in a Markovian environment, a static
or dynamic qubit undergoes an exponential relaxation to the final
statistical mixture.

It also follows from Eq.~(\ref{a12}) that , in general, the qubit's
decoherence in the steady state is incomplete and its density matrix
is not converted into a statistical mixture, $\sigma^{st}\ne {\rm
diag}\, (1/2,1/2)$ as would be the case for a point-contact
detector. Rather, complete decoherence is achieved only in the weak
coupling limit ($\alpha\to 0$) \cite{fn1} and only for the initial
conditions corresponding to Re $(ab^*)=0$. For a weak coupling, the
dependence on the qubit's initial state can be understood in a
following way. A real part of the qubit's off-diagonal
density-matrix element ${\rm Re}\, \sigma_{12}(t)$ can be written as
${\rm Re}\, \sigma_{12}(t)=(1/2)\la\Psi(t)|\hat q_+|\Psi (t)\ra$,
where $\hat q_+=d_1^\dagger d_2+d_2^\dagger d_1$. If the qubit's
levels are aligned ($E_1=E_2$), the operator $q_+$ commutes with the
total Hamiltonian, Eq.~(\ref{a1}), in the limit of $\delta\Omega\to
0$. As a result ${\rm Re}\, \sigma_{12}(t)\simeq {\rm Re}\,
\sigma_{12}(0)$. Therefore, the subspace of the qubit's states,
corresponding to ${\rm Im}\, \sigma_{12}(0)=0$ is effectively {\it
decoherence free} \cite{fn2}.

In strong coupling limit, $\alpha\to\infty$, in
Eqs.(\ref{a11}a)-(\ref{a11b}) we only need to retain the terms which
do not vanish for $|\epsilon|\to \infty$, thus compensating for the
factor  $\alpha^{-1}$ in Eq.(\ref{a10}). Accordingly, the
off-diagonal density-matrix element would disappear at all times for
all initial qubit's states, $\sigma_{12}(t) \to 0$. However, the
result is not the statistical mixture, as in the case of weak
coupling, but $\sigma (t)={\rm diag}\, (|a|^2,|b|^2)$. This
corresponds to the so-called {\it pure dephasing} for a static
($\omega_0=0$) qubit \cite{ither} whose diagonal density-matrix
elements remain constant while the off-diagonal elements vanish.

Finally, the finite size effects for a condensate with a large but
finite number of atoms are shown Fig.\ref{fig2}.
These manifest themselves as an onset of irregular oscillations
 of the qubit's density matrix (dashed lines in Fig.\ref{fig2}).
The oscillations appear at times comparable with the Rabi period of an
individual atom in the double well potential, $t\approx 2\pi \delta
\Omega$, prior to which the qubit's evolution agrees with that in
the presence of an infinite condensate. Mathematically,  the effect
occurs when the period of ever faster oscillating terms in
Eqs.(\ref{a11}a)-(\ref{a11b}) becomes comparable with the separation
between the energy levels of the condensate and Eq.(\ref{a8}) ceases
to be a valid Riemann sum for the integral (\ref{a10}). Physically,
the qubit begins to be affected by  the size of the condensate at
times of the order the Poincaire recurrence time of the latter,
i.e., when the escape of the atoms into the right well can no longer
be considered irreversible.

\section{Summary}

In summary, we have demonstrated that continuous monitoring of a
qubit by a BEC produces a slow state-selective decoherence which
obeys a power, rather that exponential, law in time (except for a
static qubit, where the decoherence is extremely fast). Although
this result was obtained in the limit $N\to\infty$, it can be
confirmed by numerical evaluations of the qubit's density-matrix,
Eq.~(\ref{a8}) for a large but finite $N$, Fig.~\ref{fig2}. It is
this non-exponential relaxation and a strong dependence on the
qubit's initial state that distinguishe the BEC model, with a single
energy level in each of the reservoirs, from the exponential
decoherence generated by a (Markovian) environment with a continuum
spectrum of available states. Common to both environments is,
however,  freezing of the qubit's internal transitions in the strong
interaction limit. This kind of Zeno effect \cite{gur1,SRES,SWEAK}
produced by the unitary evolution in the  presence of an environment
is somewhat different from its conventional prototype
\cite{zeno,zeno1} which arises from frequent observations of the
evolving system. One remarkable feature of a Markovian environment
is that the qubit's evolution under such frequent observations is
practically indistinguishable from its unitary observation-free
evolution \cite{kor1}. For a qubit-BEC hybrid system  whose behavior
is explicitly non-Markovian, we expect the two types of evolution to
be drastically different. A detailed investigation of this problem
is, however, beyond the scope of the present paper.

\begin{acknowledgements} One of us (D.S.) acknowledges the EU –
Transnational Access program, EU project \# RITA-CT-2003-506095 for
supporting his visit to the Weizmann Institute of Science, where a
part of this work has been done. We are also grateful to M.Raizen
and N. Davidson for useful discussions.
\end{acknowledgements}


\begin{thebibliography}{999}
\bibitem{symm} An asymmetry of the qubit would only lead to minor
changes in our analysis. e.g., the shift of the Gaussian in
Eq.(\ref{a10}), and not change its conclusions.
\bibitem{fn1} Note that in Eq.~(\ref{a12})
$\sqrt{\pi}\, z\exp\,(z^2)\, {\rm erfc}\, (z)\to 1$ when
$z\to\infty$.
\bibitem{fn2} A similar decoherence-free subspace can be found
for a qubit interacting with a fluctuating charge \cite{gur2}.
\end{thebibliography}
\end{document}